\documentclass{revtex4}
\usepackage{pictexwd,dcpic}
\usepackage{amsmath}
\usepackage{amsfonts}

\begin{document}

\title{Hamilton's equations in a non-associative quantum theory}
\author{Vladimir Dzhunushaliev\footnote{Senior Associate of the Abdus Salam ICTP}}
\email{dzhun@krsu.edu.kg} \affiliation{Dept. Phys. and Microel.
Engineer., Kyrgyz-Russian Slavic University, Bishkek, Kievskaya Str.
44, 720021, Kyrgyz Republic}

\date{\today}

\begin{abstract}
A new non-associative algebra for the quantization of strongly interacting fields is proposed. The full set of quantum $(\pm)$associators for the product of three operators is offered. An algorithm for the calculation of some $(\pm)$associators for the product of some four operators is offered. The possible generalization of Hamilton's equations for a non-associative quantum theory is proposed. Some arguments are given that a non-associative quantum theory can be a fundamental unifying theory. 
\end{abstract}

\pacs{}

\maketitle

\section{Introduction}

Probably the most important problem in quantum field theory is the quantization of strongly interacting fields. Ordinary canonical (anti)commutations relationships are valid for free noninteracting fields only and the algebra of interacting fields differs from the algebra of noninteracting fields. Especially such algebra is important for understanding of confinement problem in quantum chromodynamics. In this Letter we would like to show that there is a possibility to generalize the standard quantum theory on a non-associative way. We will present quantum associators (antiassociators) for the product of three and four operators, check the self-consistency of these definitions in some cases and generalize Hamilton's equations for a non-associative quantum theory. 
\par
One can point to the possible impact of non-associative algebras on: string theory \cite{Manogue:1989ey}, \cite{kugo}; dimensional reduction \cite{Manogue:1998rv}; matrix model \cite{Smolin:2001wc}; Dirac's monopole quantization procedure \cite{Grossman}, \cite{Jackiw}; non-associative quantum mechanics \cite{jordan}, \cite{okubo1995}, \cite{Dzhunushaliev:2005jb}, \cite{Dzhunushaliev:2005yd}; quark confinement \cite{Color}, \cite{gursey}; classical integrable system \cite{Dimakis:2006kp}; a unified description of continuum and discrete spacetime \cite{Nesterov:2000qb}. There are a few monographs and textbooks with mathematical definitions and physical applications of non-associative algebras in physics: \cite{okubo1995}, \cite{baez}, \cite{gursey1996}, \cite{schafer}.
\par 
In Ref's \cite{Gogberashvili:2005cp} and \cite{Gogberashvili:2005xb} Dirac's operator and Maxwell's equations are derived in the algebra of split-octonions. The results of these papers are very interesting and results in the questions: why Dirac's operator and Maxwell's equations can be formulated on the octonionic language ? It is possible that a non-associative quantum theory in some limits bring to Dirac's and Maxwell's equations that may be similar to the connection between M-theory and of all kinds of string theories. 
\par 
In conclusion we want to note that a non-associative algebra presented here is different from the non-associative algebra presented in Ref's \cite{Dzhunushaliev:2005jb}, \cite{Dzhunushaliev:2005yd}. The difference is that the anticommutators 
$\left\{ q_i, Q_i \right\}_+ = 0$ here but $\left\{ q_i, Q_i \right\}_+ = 1$ in Ref's \cite{Dzhunushaliev:2005jb}, \cite{Dzhunushaliev:2005yd}.

\section{Preliminary remarks}

In this Letter we consider operators $q_i, Q_i, i=1,2,3$ which are a non-associative operator generalization of split-octonions numbers $\tilde{q}_i, \tilde{Q}_i$ (the multiplication table of split-octonions is given in Appendix \ref{octonions}). One can say that $q_i, Q_i$ and $\tilde{q}_i, \tilde{Q}_i$ are like $q$ and $c$-Dirac numbers in the standard quantum mechanics.
\par
We now make some remarks about the calculation of the associators/antiassociators. Let the  operator $A_1$ is 
\begin{equation}
    A_1 = \left( a b \right) c
\label{sec2-10}
\end{equation}
where operators $a,b,c$ are some product of $n_{a,b,c}$ non-associative operators 
$q_i, Q_j$ and $A_2$ be the same operator but with a different arrangement of the brackets 
\begin{equation}
    A_2 = a \left( b c \right) 
\label{sec2-20}
\end{equation}
If $\tilde{a}, \tilde{b}, \tilde{c}$ are the corresponding split-octonions then $\tilde{A}_1$ and $\tilde{A}_2$ are the corresponding octonions after the replacement 
$a \rightarrow \tilde{a}$, $b \rightarrow \tilde{b}$, $c \rightarrow \tilde{c}$ and then
\begin{equation}
    \tilde A_2 = \pm \tilde A_1 .
\label{sec2-30}
\end{equation}
This permits us to define the corresponding quantum $(\pm)$associator 
\begin{equation}
  \left\{ a,b,c \right\}_\pm = A_2 + \left( \pm A_1 \right) = 
  \left( a b \right) c \pm a \left( b c \right)
\label{sec2-40}
\end{equation}
where $\left\{ a,b,c \right\}_\pm$ is the associator for $(-)$ and antiassociator for $(+)$. 
\par
We will say that the $(\pm)$associator $\left\{ a,b,c \right\}_\pm$ is $(\pm)$\textit{flexible} if 
\begin{equation}
  \left\{ a,b,c \right\}_\pm = - \left\{ c,b,a \right\}_\pm .
\label{sec2-50}
\end{equation}

\section{Quantum ($\pm$)associators for the product of three operators}

In this section we present the quantum ($\pm$)associators (which have been offered in Ref's \cite{Dzhunushaliev:2005jb}, \cite{Dzhunushaliev:2005yd}) for the product of three operators. The anticommutators are
\begin{eqnarray}
    \left\{ q_i q_j \right\}_+ &=& 0,
\label{sec2-72}\\
    \left\{ q_i Q_j \right\}_+ &=& 0,
\label{sec2-74}\\
    \left\{ Q_i Q_j \right\}_+ &=& 0
\label{sec3-10}\\
  \left\{ q_i Q_i \right\}_+ &=& 0 
\label{sec3-20}
\end{eqnarray}
and for identical indices here we have to note that the anticommutative relation \eqref{sec3-20} is different from similar relation in Ref's \cite{Dzhunushaliev:2005jb}, \cite{Dzhunushaliev:2005yd} where 
$\left\{ q_i Q_i \right\}_+ = 1$ (which are the ordinary canonical anti-commutation relationships for the creation/annihilation fermion operators). The quantum ($\pm$)associators with different indices $m \neq n, n \neq p, p \neq m $ are
\begin{eqnarray}
    \left\{ q_m,q_n,q_p \right\}_- &=& 
    \left( q_m q_n \right) q_p - q_m\left( q_n q_p \right) = 0 ,
\label{sec3-30}\\
    \left\{ Q_m,Q_n,Q_p \right\}_+ &=& 
    \left( Q_mQ_n \right) Q_p + Q_m\left( Q_n Q_p \right) =
    \epsilon_{mnp} \mathcal{H}_{3,1} ,
\label{sec3-40}\\
    \left\{ q_m,Q_n,q_p \right\}_+ &=& 
    \left( q_m Q_n \right) q_p + q_m\left( Q_n q_p \right) =
    \epsilon_{mnp} \mathcal{H}_{3,2} ,
\label{sec3-50}\\
    \left\{ Q_m,q_n,Q_p \right\}_- &=& 
    \left( Q_m q_n \right) Q_p - Q_m\left( q_n Q_p \right) =
    \epsilon_{mnp} \mathcal{H}_{3,3} ,
\label{sec3-60}\\
    \left\{ q_m,q_n,Q_p \right\}_- &=& 
    \left( q_m q_n \right) Q_p - q_m\left( q_n Q_p \right) =
    \epsilon_{mnp} \mathcal{H}_{3,4} ,
\label{sec3-70}\\
    \left\{ Q_m,q_n,q_p \right\}_- &=& 
    \left( Q_m q_n \right) q_p - Q_m\left( q_n q_p \right) =
    \epsilon_{mnp} \mathcal{H}_{3,5} ,
\label{sec3-80}\\
    \left\{ q_m,Q_n,Q_p \right\}_- &=& 
    \left( q_m Q_n \right) Q_p - q_m\left( Q_n Q_p \right) =
    \epsilon_{mnp} \mathcal{H}_{3,6} ,
\label{sec3-90}\\
    \left\{ Q_m,Q_n,q_p \right\}_- &=& 
    \left( Q_m Q_n \right) q_p - Q_m\left( Q_n q_p \right) =
    \epsilon_{mnp} \mathcal{H}_{3,7}
\label{sec3-100}
\end{eqnarray}
where $\mathcal{H}_{3,i}$ are some operators and in contrast with the previous papers \cite{Dzhunushaliev:2005jb} \cite{Dzhunushaliev:2005yd} we assume that these operators can be new operators not the linear combination of $q, Q$ and $q Q$; Eq. \eqref{sec3-30} means that the quaternionic - like subalgebra spanned on $q_1, q_2, q_3$ is the associative algebra. It is easy to see that $(\pm)$associators \eqref{sec3-40}-\eqref{sec3-100} are $(\pm)$flexible. The quantum antiassociators for the product of three operators, such as $q(Qq)$ or $Q(qQ)$, having 
two different indices $m \neq n$ are
\begin{eqnarray}
    \left\{ q_m,Q_n,q_n \right\}_+ &=& 
    q_m\left( Q_n q_n \right) + \left( q_m Q_n \right) q_n = \mathcal{H}_{3,8}(m,n) ,
\label{sec3-120}\\
    \left\{ q_n,Q_n,q_m \right\}_+ &=& 
    q_n\left( Q_n q_m \right) + \left( q_n Q_n \right) q_m = \mathcal{H}_{3,9}(m,n) ,
\label{sec3-130}\\
    \left\{ Q_m,q_n,Q_n \right\}_+ &=& 
    Q_m\left( q_n Q_n \right) + \left( Q_m q_n \right) Q_n = \mathcal{H}_{3,10}(m,n) ,
\label{sec3-140}\\
    \left\{ Q_n,q_n,Q_m \right\}_+ &=& 
    Q_n\left( q_n Q_m \right) + \left( Q_n q_n \right) Q_m = \mathcal{H}_{3,11}(m,n) .
\label{sec3-150}
\end{eqnarray}
The quantum associators, such as $q(QQ)$ or $Q(qq)$, and with two different indices $m \neq n$ are
\begin{eqnarray}
    \left\{ q_m, Q_m, Q_n \right\}_- &=& 
    \left( q_m Q_m \right) Q_n - q_m\left( Q_m Q_n \right) = \mathcal{H}_{3,12}(m,n) ,
\label{sec3-160}\\
    \left\{ q_m, Q_n, Q_m \right\}_- &=& 
   \left( q_m Q_n \right) Q_m - q_m\left( Q_n Q_m \right) = \mathcal{H}_{3,13}(m,n) ,
\label{sec3-170}\\
    \left\{ Q_n, Q_m, q_m \right\}_- &=& 
    \left( Q_n Q_m \right) q_m - Q_n\left( Q_m q_m \right) = \mathcal{H}_{3,14}(m,n) ,
\label{sec3-180}\\
    \left\{ Q_m, Q_n, q_m \right\}_- &=& 
    \left( Q_m Q_n \right) q_m - Q_m\left( Q_n q_m \right) = \mathcal{H}_{3,15}(m,n) ,
\label{sec3-190}\\
    \left\{ Q_m, q_m, q_n \right\}_- &=& 
    \left( Q_m q_m \right) q_n - Q_m\left( q_m q_n \right) = \mathcal{H}_{3,16}(m,n) ,
\label{sec3-200}\\
    \left\{ Q_m, q_n, q_m \right\}_- &=& 
    \left( Q_m q_n \right) q_m - Q_m\left( q_n q_m \right) = \mathcal{H}_{3,17}(m,n) ,
\label{sec3-210}\\
    \left\{ q_n, q_m, Q_m \right\}_- &=& 
    \left( q_n q_m \right) Q_m - q_n\left( q_m Q_m \right) = \mathcal{H}_{3,18}(m,n) ,
\label{sec3-220}\\
    \left\{ q_m, q_n, Q_m \right\}_- &=& 
    \left( q_m q_n \right) Q_m - q_m\left( q_n Q_m \right) = \mathcal{H}_{3,19}(m,n) 
\label{sec3-230}
\end{eqnarray}
where $\mathcal{H}_{3,i}(m,n)$ are for the time undefined operators. The alternativity properties are
\begin{eqnarray}
    \left\{ q_n, q_n, Q_m \right\}_- &=& 0,
\label{sec3-240}\\
    \left\{ q_n, Q_m, q_n \right\}_- &=& 0,
\label{sec3-250}\\
    \left\{ Q_m, q_n, q_n \right\}_- &=& 0.
\label{sec3-260}
\end{eqnarray}

\section{Self - consistency of quantum ($\pm$)associators for the product of three operators}

In this section we will check the self-consistency of the $(\pm)$associators for the product of three operators. For this we will permute the first and third factors in the product $a(bc)$ following to the commutative diagram 
\begin{equation}
\begindc{\commdiag}[3]
\obj(1,5){$a \left( bc \right)$}
\mor(4,5)(15,11){}
\mor(4,5)(15,-1){}

\obj(17,10){$a \left( cb \right)$}
\obj(32,10){$\left(ac \right) b$}
\obj(47,10){$\left(ca \right) b$}
\obj(62,10){$c \left(ab \right)$}
\mor(19,10)(30,10){}
\mor(34,10)(45,10){}
\mor(49,10)(60,10){}

\obj(17,0){$\left( ab \right) c$}
\obj(32,0){$\left( ba \right) c$}
\obj(47,0){$b \left( ac \right)$}
\obj(62,0){$b \left(ca \right)$}
\obj(77,0){$\left( bc \right) a$}
\obj(92,0){$\left( cb \right) a$}
\mor(19,0)(30,0){}
\mor(34,0)(45,0){}
\mor(49,0)(60,0){}
\mor(64,0)(75,0){}
\mor(79,0)(90,0){}

\obj(110,5){$c \left( ba \right)$}
\mor(95,0)(107,5){}
\mor(65,11)(107,5){}
\enddc
\label{sec3a-10}
\end{equation}
The quantum associators \eqref{sec3-40}-\eqref{sec3-100} are the same as in Ref.~\cite{Dzhunushaliev:2005jb} and consequently the verification of the self-consistency is the same. 
\par 
The calculations with two equal indices give us two possibilities. The first one is 
\begin{equation}
	q_1 \left( Q_1 Q_2 \right) = Q_2 \left( Q_1 q_1 \right) + 
	\mathcal H_{3,13}\left( 1,2 \right) + \mathcal H_{3,10}\left( 2,1 \right)
\label{sec3a-20}
\end{equation}
here we use the upper row of the diagram \eqref{sec3a-10}. The lower row gives us 
\begin{equation}
	q_1 \left( Q_1 Q_2 \right) = Q_2 \left( Q_1 q_1 \right) - 
	\mathcal H_{3,12}\left( 1,2 \right) - \mathcal H_{3,11}\left( 2,1 \right) + 
	\mathcal H_{3,15}\left( 1,2 \right) + \mathcal H_{3,14}\left( 1,2 \right).
\label{sec3a-30}
\end{equation}
Comparing the rhs of Eq's \eqref{sec3a-20} and \eqref{sec3a-30} one can assume that 
\begin{eqnarray}
  \mathcal H_{3,10}\left( 2,1 \right) &=& - \mathcal H_{3,11}\left( 2,1 \right) ,
\label{sec3a-40}\\
  \mathcal H_{3,12}\left( 1,2 \right) + \mathcal H_{3,13}\left( 1,2 \right) 
  &=& 
  \mathcal H_{3,14}\left( 1,2 \right) + \mathcal H_{3,15}\left( 1,2 \right) .
\label{sec3a-50}
\end{eqnarray}
The pair indices $(1,2)$ in Eq's \eqref{sec3a-40} \eqref{sec3a-50} can be replaced by any other pair indices $(i,j)$. 
\par 
Eq. \eqref{sec3a-40} tells us that the associators \eqref{sec3-140} \eqref{sec3-150} are (+)flexible 
\begin{equation}
	\left\{ Q_m,q_n,Q_n \right\}_+ = - 
	\left\{ Q_n,q_n,Q_m \right\}_+ 
\label{sec3a-60}
\end{equation}
For the second case the upper row of the diagram Eq. \eqref{sec3a-10} gives us 
\begin{equation}
	Q_1 \left( q_1 q_2 \right) = q_2 \left( q_1 Q_1 \right) + 
	\mathcal H_{3,8}\left( 2,1 \right) + \mathcal H_{3,17}\left( 1,2 \right) .
\label{sec3a-70}
\end{equation}
The lower row gives us 
\begin{equation}
	Q_1 \left( q_1 q_2 \right) = q_2 \left( q_1 Q_1 \right) + 
	\mathcal H_{3,18}\left( 1,2 \right) + \mathcal H_{3,19}\left( 1,2 \right) - 
	\mathcal H_{3,16}\left( 1,2 \right) - \mathcal H_{3,9}\left( 2,1 \right).
\label{sec3a-80}
\end{equation}
Comparing the rhs of Eq's \eqref{sec3a-70} and \eqref{sec3a-80} one can assume that 
\begin{eqnarray}
  \mathcal H_{3,8}\left( 2,1 \right) &=& - \mathcal H_{3,9}\left( 2,1 \right) ,
\label{sec3a-90}\\
  \mathcal H_{3,16}\left( 1,2 \right) + \mathcal H_{3,17}\left( 1,2 \right) 
  &=& 
  \mathcal H_{3,18}\left( 1,2 \right) + \mathcal H_{3,19}\left( 1,2 \right) .
\label{sec3a-100}
\end{eqnarray}
The pair indices $(1,2)$ in Eq's \eqref{sec3a-70} \eqref{sec3a-80} can be replaced by any other pair indices $(i,j)$. 
\par
Eq. \eqref{sec3a-90} tells us that the associators \eqref{sec3-120} \eqref{sec3-130} are (+)flexible 
\begin{equation}
	\left\{ q_m,Q_n,q_n \right\}_+ = - 
	\left\{ q_n,Q_n,q_m \right\}_+ 
\label{sec3a-110}
\end{equation}

\section{Quantum ($\pm$)associators for the product of four operators}

The basic idea for this section is to define quantum ($\pm$)associators for the product of four operators using ($\pm$)associators for the product of three operators. For this we assume the following algorithms 
\begin{eqnarray}
  \left\{ ab, c ,d \right\}_\pm &\equiv& 
  \left( \left( ab \right) c \right) d \pm \left( ab \right) \left( cd \right) = 
  a \left\{ b, c ,d \right\}_\pm + \left\{ a, c ,d \right\}_\pm b ,
\label{sec4-10}\\
  \left\{ a, b, cd \right\}_\pm &\equiv& 
  \left( ab \right) \left( cd \right) \pm a \left( b \left( cd \right) \right) = 
  \left\{ a, b, c \right\}_\pm d + c \left\{ a, b ,d \right\}_\pm 
\label{sec4-20}
\end{eqnarray}
where $a,b,c,d$ are the octonion-like operators $q_i, Q_i$. 
\par 
Now we will check the rules \eqref{sec4-10} and \eqref{sec4-20} for some products using the following commutative diagram 
\begin{equation}
\begindc{\commdiag}[3]
\obj(0,5){$\left( ab \right)\left( cd \right)$}
\mor(4,5)(15,11){}
\mor(4,5)(15,-1){}

\obj(20,10){$\left(\left( ab \right) c \right)d$}
\obj(41,10){$\left( a \left( bc \right)\right)d$}
\obj(62,10){$\left( a \left( cb \right)\right)d$}
\obj(83,10){$\left(\left( ac \right) b \right)d$}
\mor(25,10)(37,10){}
\mor(46,10)(58,10){}
\mor(67,10)(79,10){}

\obj(20,0){$a \left( b \left( cd \right)\right)$}
\obj(41,0){$a \left(\left( bc \right) d \right)$}
\obj(62,0){$a \left(\left( cb \right) d \right)$}
\obj(83,0){$a \left( c \left( bd \right)\right)$}
\mor(25,0)(37,0){}
\mor(46,0)(58,0){}
\mor(67,0)(79,0){}

\obj(104,5){$\left( ac \right)\left( bd \right)$}
\mor(88,-1)(100,6){}
\mor(88,11)(100,4){}
\enddc
\label{sec4-30}
\end{equation}
here we permute the second and third factors $b,c$. 
\par
Now we will calculate a few examples. Using the commutative diagram \eqref{sec4-30} we have for the product $\left( q_1 q_2 \right) \left( q_3 Q_3 \right)$
\begin{eqnarray}
	\left( q_1 q_2 \right) \left( q_3 Q_3 \right) &=& 
	- \left( q_1 q_3 \right) \left( q_2 Q_3 \right) - 
	q_1 \left(
		\mathcal H_{3,18} \left( 3,2 \right) + \mathcal H_{3,19} \left( 3,2 \right)
	\right) - 
	\mathcal H_{3,18} \left( 3,1 \right) q_2 - 
	\mathcal H_{3,4} q_3 = 
\label{sec4-40}\\
	&&- \left( q_1 q_3 \right)\left( q_2 Q_3 \right) - 
	q_1 \left(
		\mathcal H_{3,18} \left( 3,2 \right) + \mathcal H_{3,19} \left( 3,2 \right)
	\right) + 
	q_2 \mathcal H_{3,18} \left( 3,1 \right) + 
	q_3 \mathcal H_{3,4}
\label{sec4-50}
\end{eqnarray}
Eq's \eqref{sec4-40} and \eqref{sec4-50} are deduced using upper and lower rows of commutative diagram \eqref{sec4-30} correspondingly (for details of these calculations,  see appendix \ref{examples}). Comparing Eq's \eqref{sec4-40} and \eqref{sec4-50} we have 
\begin{eqnarray}
	q_2 \mathcal H_{3,18} \left( 3,1 \right) &=& - 
	\mathcal H_{3,18} \left( 3,1 \right) q_2, 
\label{sec4-60}\\
	q_3 \mathcal H_{3,4} &=& - \mathcal H_{3,4} q_3. 
\label{sec4-70}
\end{eqnarray}
It is evidently that Eq's \eqref{sec4-40} and \eqref{sec4-50} are correct for any permutations of indices $(1,2,3) \rightarrow (i,j,k)$. It means that the operator 
$\mathcal H_{3,4}$ can not include the operators $q_{1,2,3}$ as the terms. 
\par
But not all is so simple here. If we apply the procedure \eqref{sec4-30} to the $(-)$associator $\left\{ q_1Q_1, q_3, Q_3 \right\}_-$ we will see that the corresponding diagram is not commutative. The thorough analysis shows that the reason of this difference is that $\left\{ q_1Q_1, q_3, Q_3 \right\}_-$ $(-)$associator has two pairs $(q_1 Q_1)$ and $(q_3 Q_3)$. In Ref. \cite{Dzhunushaliev:2005jb} the notion of white (colorless) operators is introduced: 

the white operators, $\mu_i, i=1,2,3$ are
\begin{equation}
    \mu_i = q_i Q_i ~
\label{sec4-80}
\end{equation}
here $i$ is not summed over. Thus we can assume that \eqref{sec4-10} and \eqref{sec4-20} algorithms are valid only for four $a,b,c,d$ which is not any combination of four $q_i, Q_i, q_j, Q_j$. In this case we have to assume 
\begin{eqnarray}
	\left\{ q_i Q_i, q_j, Q_j \right\}_{\pm} &=& \mathcal H_{4,1} , 
\label{sec4-90}\\
	\left\{ q_i, Q_i, q_j Q_j \right\}_{\pm} &=& \mathcal H_{4,2} 
\label{sec4-100}
\end{eqnarray}
are new operators. 
\par
The definitions \eqref{sec4-10} \eqref{sec4-20} one can use for the definitions of other ($\pm$)associators. For example, one can use the commutative diagram
\begin{equation}
\begindc{\commdiag}[3]
\obj(14,11){$\left( \left( ab \right) c \right) d$}
\obj(39,11){$\left( ab \right) \left( cd \right)$}
\obj(64,11){$a \left( b \left( cd \right) \right)$}
\obj(14,1){$\left( a \left( bc \right) \right) d$}
\obj(64,1){$a \left( \left( bc \right) d\right)$}
\mor(18,11)(35,11){\footnotesize{Eq. \eqref{sec4-10}}}
\mor(43,11)(60,11){\footnotesize{Eq. \eqref{sec4-20}}}
\mor(18,1)(60,1){\footnotesize{Eq. \eqref{sec4-80}}}
\mor(14,11)(14,1){}
\mor(64,11)(64,1){}
\enddc
\label{sec4-110}
\end{equation}
to calculate 
\begin{equation}
	\left\{
		a, bc, d
	\right\}_\pm = 
	\left( a \left( bc \right) \right) d \pm a \left( \left( bc \right) d\right)
\label{sec4-120}
\end{equation}

\section{Hamilton's equations}
\label{hamilton_eq}

The essential ingredient of any quantum theory is dynamical equations. Ordinary they are  Hamilton's equation(s) 
\begin{equation}
	\frac{d \widehat{L}}{dt} = i \left[ \widehat{H}, \widehat{L} \right]
\label{sec6-10}
\end{equation}
where $\widehat{L}$ is some operator, $\widehat{H}$ is the Hamiltonian. Eq. \eqref{sec6-10} leads to the result that the commutator $[,]$ should be a derivative 
\begin{equation}
	\left[ \widehat{A}, \widehat{B} \widehat{C} \right] = 
	B \left[ \widehat{A}, \widehat{C} \right] + 
	\left[ \widehat{A}, \widehat{B} \right] \widehat{C} .
\label{sec6-20}
\end{equation}
The question here is how we can generalize Hamilton's equation(s) into a non-associative operator algebra. The well known result is the Myung's Theorem \cite{myung}: 
\par 
A necessary and sufficient condition that an algebra $A$ over the field $F$ of characteristic not two be flexible Lie-admissible is to have Eq. \eqref{sec6-20}. This is precisely the derivative law 
\begin{equation}
	\frac{d}{dt} \left(
		\widehat{A} \widehat{B}
	\right) = \widehat{A} \frac{d \widehat{B}}{dt} + 
	\frac{d \widehat{A}}{dt} \widehat{B} 
\label{sec6-30}
\end{equation}
so that the consistent quantization requires the underlying operator algebra to be flexible Lie-admissible. Only in this case Hamilton's equation(s) \eqref{sec6-10} will be self-consistent. 
\par 
An algebra $A$ is said to be flexible if 
\begin{equation}
	\left\{ x,y,z \right\}_- = - \left\{ z,y,x \right\}_- 
\label{sec6-40}
\end{equation}
where $\left\{ , ,\right\}_-$ is the associator; $x,y,z \in A$. An algebra $A$ is called Lie-admissible if the Jacobi identity is valid
\begin{equation}
	\left[ x, \left[ y,z \right]\right] + \left[ z, \left[ x,y \right]\right] + 
	\left[ y, \left[ z,x \right]\right] = 0.
\label{sec6-50}
\end{equation}
Let us note that we use \textit{(-)associators} and \textit{commutators} in \eqref{sec6-40} \eqref{sec6-50} equations correspondingly. For probable generalization of this procedure we can: (a) change the commutator in Hamilton's equation \eqref{sec6-10} on $(\pm)$associator, (b) introduce associators and antiassociators on equal rights. 
\par 
Thus the conclusion is to change rhs of Hamilton's equation \eqref{sec6-10} for a non-associative quantum theory by the following way 
\begin{equation}
	\frac{d a}{dt} = \left\{
		a,c,d
	\right\}_\pm = \left(ac\right) d \pm a \left( cd \right)
\label{sec6-70}
\end{equation}
where $\left\{ , , \right\}_\pm$ is $(\pm)$associator; $a,b,c$ are non-associative operators and $(cd)$ is \emph{a non-associative Hamiltonian}. With the definitions \eqref{sec4-10} and \eqref{sec4-20} the derivation law 
\begin{equation}
	\frac{d (ab)}{dt} = 
	a \frac{d b}{dt} + \frac{d b}{dt}a = 
	\left\{
		ab,c, d
	\right\}_\pm = 
	a\left\{
		b,c, d
	\right\}_\pm + 
	\left\{
		a,c, d
	\right\}_\pm b
\label{sec6-80}
\end{equation}
is valid at least for some four $a,b,c,d$. According to remark at the end of section \ref{hamilton_eq} a non-associative Hamilton's equation \eqref{sec6-70} is valid only for the case if $(ab)$ and $(cd)$ operators are not the white operators simultaneously. If $(ab)$ and $(cd)$ are the colorless operators simultaneously the problem of the definition $\frac{d(ab)}{dt}$ with a white Hamiltonian remains an open problem. But we have to do the following remark. A non-associative algebra of quantum operators considered here should have associative subalgebras, for example it is the subalgebra spanned on quaternion-like operators $q_{1,2,3}$. One can assume that the white operators form associative subalgebras, so that 
\begin{equation}
	\frac{d (q_i Q_i)}{dt} = \left[ q_i Q_i, q_j Q_j \right]
\label{sec6-90}
\end{equation}
where $[,]$ is ordinary commutator.

\par 
On this way there is one big problem: it is necessary to proof the self-consistence of the definitions \eqref{sec4-10} and \eqref{sec4-20} for all four $(a,b,c,d)$ and for the product of $n>4$ operators. 

\section{Summary and conclusions}

In this Letter we offer a new non-associative algebra which can be considered as an algebra of strongly interacting fields. The values of some quantum $(\pm)$associators is offered. The self-consistency of the algebra on the level of the product of three operators and for some products of four operators is checked. The prior analysis shows us that the operators $\mathcal H_{3,1-7}$ are new operators which are not the combination of $q_i, Q_i$ operators. Another interesting property of the considered algebra is that $(\pm)$associators $\mathcal H_{3,1-11}$ are $(\pm)$flexible. For the definition of the product of $n>4$ operators it is necessary either to consider $(\pm)$associators for the product of $n>4$ operators following to an inductive algorithm (\eqref{sec4-10} \eqref{sec4-20} for example) or to do some new assumptions about these $(\pm)$associators. One can assume that so-called \emph{an invisibility principle} \cite{Dzhunushaliev:2005yd} is valid. 
\par
This principle tells us: \textit{in a calculation with rearrangements of brackets in a product the white operator is invisible. This implies that if there is a colorless operator on the lhs or rhs then the corresponding quantum $(\pm)$associator does not have any operators which were in the colorless operator.}
\par 
The white operators are operators where the change of octonion-like operators $q, Q$ on the split-octonions bring to the split-octonion $I$ (for the definition of the split-octonion $I$ see Table \ref{octonions}). For example the operators $q_i Q_i$ are white operators. In this case, for example, 
\begin{equation}
    \left\{ q_m,Q_n,q_n \right\}_+ = 
    q_m\left( Q_n q_n \right) + \left( q_m Q_n \right) q_n = 
    a q_m
\label{sec7-10}
\end{equation}
where $a$ is some real number. 
\par 
The most intriguing problem for physicists in this consideration is: how one can apply a non-associative algebras in physics ? The authors point of view is that if in a non-associative algebra there are associative algebras then only the elements of these associative subalgebras have physical sense. It means that only these associative elements are physical observables. The non-associative elements are not physical observables. 
\par 
If non-associative quantum theory has such associative subalgebras as u(1), su(2), su(3), so(1,3) or even any supersymmetry  algebra then such theory can be considered as a fundamental theory which unifies gauge theories, gravity and even supersymmetrical theories. 

\appendix
\section{The split-octonions}
\label{octonions}

Let split-octonion numbers are designeted as $\tilde{q}_i$ and $\tilde{Q}_j$. Here we present the multiplication rule of the split-octonions numbers $\tilde{q}_i, \tilde{Q}_j$  and $I$. 
\begin{table}[h]
\begin{tabular}{|c|c|c|c|c|c|c|c|c|}                                                
\hline
&$\tilde{q}_1$ & $\tilde{q}_2$ & $\tilde{q}_3$ & $\tilde{Q}_1$  & $\tilde{Q}_2$ & $\tilde{Q}_3$ & $I$         \\ 
\hline 
$\tilde{q}_1$& $-1  $ & $\tilde{q}_3$  & $-\tilde{q}_2$  & $-I$ & $\tilde{Q}_3$  & $-\tilde{Q}_2$ & $\tilde{Q}_1$ \\ 
\hline
$\tilde{q}_2$ & $-\tilde{q}_3$ & $-1$   & $\tilde{q}_1$  & $-\tilde{Q}_3$  & $-I$ & $\tilde{Q}_1$  & $\tilde{Q}_2$    \\ 
\hline
$\tilde{q}_3$ & $\tilde{q}_2$ & $-\tilde{q}_1$ & $-1$   & $\tilde{Q}_2$  & $-\tilde{Q}_1$  & $-I$ & $\tilde{Q}_3$    \\ 
\hline
$\tilde{Q}_1$ & $I $ & $\tilde{Q}_3$ & $-\tilde{Q}_2$ & $ 1$   & $-\tilde{q}_3$  & $\tilde{q}_2$  & $\tilde{q}_1$    \\ 
\hline
$\tilde{Q}_2$ & $-\tilde{Q}_3$ & $I$  & $\tilde{Q}_1$ & $\tilde{q}_3$ & $ 1$   & $-\tilde{q}_1$  & $\tilde{q}_2$    \\ 
\hline
$\tilde{Q}_3$ & $\tilde{Q}_2 $ & $-\tilde{Q}_1$ & $I$  & $-\tilde{q}_2$ & $\tilde{q}_1$ & $ 1$   & $\tilde{q}_3$    \\ 
\hline
$I$ & $-\tilde{Q}_1 $ & $-\tilde{Q}_2$  & $-\tilde{Q}_3$ & $-\tilde{q}_1$  & $-\tilde{q}_2$ & $-\tilde{q}_3$ & $ 1$    \\ 
\hline
\end{tabular}
\caption{The split-octonions multiplication table.} 
\label{oct}
\end{table}

\section{A few examples for the calculation of four $(\pm)$associators}
\label{examples}

At first we will calculate the following permutation 
\begin{equation}
\begin{split}
	\left( q_1 q_2 \right) \left( q_3 Q_3 \right) \rightarrow 
	\left( q_1 q_3 \right) \left( q_2 Q_3 \right) .
\label{appb-10}
\end{split}
\end{equation}
In order to calculate this permutation we start from the definition \eqref{sec4-10}
\begin{equation}
\begin{split}
	\left\{ q_1 q_2, q_3, Q_3 \right\}_- = & 
	\left(\left( q_1 q_2 \right) q_3 \right) Q_3 - 
	\left( q_1 q_2 \right) \left( q_3 Q_3 \right) = 
	q_1 \left\{ q_2, q_3, Q_3 \right\}_- + \left\{ q_1 , q_3, Q_3 \right\}_- q_2 = 
	\\
	& q_1 \mathcal H_{3,18} \left( 3,2 \right) + 
	\mathcal H_{3,18} \left( 3,1 \right) q_2 .
\label{appb-20}
\end{split}
\end{equation}
Whereupon we have 
\begin{equation}
\begin{split}
	\left( q_1 q_2 \right) \left( q_3 Q_3 \right) = & 
	\left(\left( q_1 q_2 \right) q_3 \right) Q_3 - 
	q_1 \mathcal H_{3,18} \left( 3,2 \right) - 
	\mathcal H_{3,18} \left( 3,1 \right) q_2 = 
	\\
	&
	- \left(\left( q_1 q_3 \right) q_2 \right) Q_3 - 
	q_1 \mathcal H_{3,18} \left( 3,2 \right) - 
	\mathcal H_{3,18} \left( 3,1 \right) q_2 .
\label{appb-30}
\end{split}
\end{equation}
Eq. \eqref{appb-30} follows form the associativity of quaternions $q_1, q_2 , q_3$. Now we use the following (-)associator 
\begin{equation}
\begin{split}
	\left\{ q_1 q_3, q_2, Q_3 \right\}_- = & 
	\left(\left( q_1 q_3 \right) q_2 \right) Q_3 - 
	\left( q_1 q_3 \right) \left( q_2 Q_3 \right) = 
	q_1 \left\{ q_3, q_2, Q_3 \right\}_- + \left\{ q_1 , q_2, Q_3 \right\}_- q_3 = 
	\\
	& q_1 \mathcal H_{3,19} \left( 3,2 \right) + 
	\mathcal H_{3,4} q_3 .
\label{appb-40}
\end{split}
\end{equation}
Eq. \eqref{appb-40} leads to 
\begin{equation}
	- \left(\left( q_1 q_3 \right) q_2 \right) Q_3 = 
	-\left( q_1 q_3 \right) \left( q_2 Q_3 \right) - 
	q_1 \mathcal H_{3,19} \left( 3,2 \right) - 
	\mathcal H_{3,4} q_3 .
\label{appb-50}
\end{equation}
Combining Eq's \eqref{appb-30} and \eqref{appb-50} we have 
\begin{equation}
\begin{split}
	\left( q_1 q_2 \right) \left( q_3 Q_3 \right) = & 
	- \left( q_1 q_3 \right) \left( q_2 Q_3 \right) - 
	q_1 \left(
		\mathcal H_{3,18} \left( 3,2 \right) + \mathcal H_{3,19} \left( 3,2 \right)
	\right) - 
	\mathcal H_{3,18} \left( 3,1 \right) q_2 - 
	\mathcal H_{3,4} q_3 .
\label{appb-60}
\end{split}
\end{equation}
Eq. \eqref{appb-60} is calculated using the upper row of commutative diagram \eqref{sec4-30}. 
\par 
Now we would like to calculate the same permutation using the lower row of commutative diagram \eqref{sec4-30}. For this we will use the definition \eqref{sec4-20} 
\begin{equation}
	\left\{ q_1, q_2, q_3 Q_3 \right\}_- = 
	\left( q_1 q_2 \right) \left( q_3 Q_3 \right) - 
	q_1 \left( q_2 \left( q_3 Q_3 \right)\right) = 
	q_3 \left\{ q_1, q_2 , Q_3 \right\}_- + 
	\left\{ q_1, q_2, q_3 \right\}_-Q_3 = 
	q_3 \mathcal H_{3,4} .
\label{appb-70}
\end{equation}
From Eq. \eqref{appb-70} we have 
\begin{equation}
\begin{split}
	\left( q_1 q_2 \right) \left( q_3 Q_3 \right) = &
	q_1 \left( q_2 \left( q_3 Q_3 \right)\right) + q_3 \mathcal H_{3,4} = 
	q_1 \left(\left( q_2 q_3 \right) Q_3 \right) - 
	q_1 \mathcal H_{3,18} \left( 3,2 \right) + q_3 \mathcal H_{3,4} = 
	\\
	&
	- q_1 \left(\left( q_3 q_2 \right) Q_3 \right) - 
	q_1 \mathcal H_{3,18} \left( 3,2 \right) + q_3 \mathcal H_{3,4} = 
	\\
	& 
	-q_1 \left( q_3 \left(q_2 Q_3 \right)\right) - 
	q_1 \left( 
		\mathcal H_{3,18} \left( 3,2 \right) + \mathcal H_{3,19} \left( 3,2 \right)
	\right) + q_3 \mathcal H_{3,4} .
\label{appb-80}
\end{split}
\end{equation}
To calculate 
$q_1 \left( q_3 \left( q_2 Q_3 \right)\right) \rightarrow 
\left( q_1 q_3 \right) \left( q_2 Q_3 \right)$ we use the definition \eqref{sec4-20} 
\begin{equation}
	\left\{ q_1, q_3, q_2 Q_3 \right\}_- = 
	\left( q_1 q_3 \right)\left( q_2 Q_3 \right) - 
	q_1 \left( q_3 \left( q_2 Q_3 \right)\right) = 
	q_2 \left\{ q_1, q_3, Q_3 \right\}_- + 
	\left\{ q_1, q_3, q_2 \right\}_- Q_3 = 
	q_2 \mathcal H_{3,18} \left( 3,1 \right) .
\label{appb-90}
\end{equation}
Whereupon we have 
\begin{equation}
	- q_1 \left( q_3 \left( q_2 Q_3 \right)\right) = 
	- \left( q_1 q_3 \right)\left( q_2 Q_3 \right) + 
	q_2 \mathcal H_{3,18} \left( 3,1 \right) .
\label{appb-100}
\end{equation}
Combining Eq's \eqref{appb-80} and \eqref{appb-100} we have 
\begin{equation}
\begin{split}
	\left( q_1 q_2 \right) \left( q_3 Q_3 \right) = & 
	- \left( q_1 q_3 \right)\left( q_2 Q_3 \right) - 
	q_1 \left(
		\mathcal H_{3,18} \left( 3,2 \right) + \mathcal H_{3,19} \left( 3,2 \right)
	\right) + 
	q_2 \mathcal H_{3,18} \left( 3,1 \right) + 
	q_3 \mathcal H_{3,4} .
\label{appb-110}
\end{split}
\end{equation}
The similar calculation for the permutation 
$\left( Q_1 Q_2 \right)\left( q_3 Q_3 \right) \rightarrow 
\left( Q_1 q_3 \right)\left( Q_2 Q_3 \right)$ give us the following result 
\begin{eqnarray}
	&&\left( Q_1 Q_2 \right)\left( q_3 Q_3 \right) =
\nonumber \\
	&&-\left( Q_1 q_3 \right)\left( Q_2 Q_3 \right) + 
	\left( -\mathcal H_{3} + \mathcal H_{7} \right) Q_3 + 
	Q_1 \left(
		-\mathcal H_{3,10} \left( 2,3 \right) - \mathcal H_{3,13} \left( 3,2 \right)
	\right) - 
	\mathcal H_{3,10} \left( 1,3 \right) Q_2 - 
	\mathcal H_{1} q_3 = 
\label{appb-120}\\
	&&-\left( Q_1 q_3 \right)\left( Q_2 Q_3 \right) + 
	\left( -\mathcal H_{3} + \mathcal H_{7} \right) Q_3 + 
	Q_1 \left(
		-\mathcal H_{3,10} \left( 2,3 \right) - \mathcal H_{3,13} \left( 3,2 \right)
	\right) + 
	Q_2 \mathcal H_{3,10} \left( 1,3 \right) + 
	q_3 \mathcal H_{1} .
\label{appb-130}
\end{eqnarray}
From what we see that 
\begin{eqnarray}
	\mathcal H_{3,10} \left( 1,3 \right) Q_2 &=& 
	- Q_2 \mathcal H_{3,10} \left( 1,3 \right) ,
\label{appb-140}\\
	\mathcal H_{1} q_3 &=& 
	- q_3 \mathcal H_{1} .
\label{appb-150}
\end{eqnarray}

\end{document}